\documentstyle[preprint,aps,eqsecnum]{revtex}
\begin{document}
\draft

\title{Twist Four Longitudinal 
	Structure Function in Light-Front QCD}

\author{A. Harindranath\thanks{e-mail: hari@tnp.saha.ernet.in}, 
Rajen Kundu\thanks{e-mail: rajen@tnp.saha.ernet.in}, and Asmita
Mukherjee\thanks{e-mail: asmita@tnp.saha.ernet.in} \\
Saha Institute of Nuclear Physics, 1/AF, Bidhan Nagar, 
	Calcutta 700064 India \\
James P. Vary\thanks{e-mail: jvary@iastate.edu} \\
 International Institute of Theoretical and Applied Physics \\
     Iowa State University, Ames, IA 50011, U.S.A. \\
and \\
 Physics Department,
     Iowa State University, Ames, IA 50011, U.S.A.}

\date{April 30, 1998}

\maketitle
\begin{abstract}
To resolve various outstanding issues associated with the twist four
longitudinal structure function ${F_L^{\tau=4}(x)}$ we perform an analysis 
based on the BJL expansion for the forward virtual photon-hadron Compton 
scattering amplitude and equal (light-front) time current algebra. Using the 
Fock space expansion for states and operators, we evaluate the twist four 
longitudinal structure function for dressed quark and gluon targets 
in perturbation theory. With the help of a new sum rule which we have derived 
recently we show that the quadratic and logarithmic divergences generated in 
the bare theory are related to corresponding mass shifts in old-fashioned 
light-front perturbation theory. We present numerical results for the $F_2$ 
and $F_L$ structure functions for the meson in two-dimensional QCD in the one 
pair approximation. We discuss the relevance of our results for the problem of 
the partitioning of hadron mass in QCD.    
\end{abstract}
\section{Introduction}
An important problem in applying QCD to deep inelastic scattering is the
existence of power corrections to scaling, more commonly known as higher
twist effects. They are
essential for making precision tests of QCD. From the early days of the 
establishment of QCD as the
underlying theory of strong interactions, the importance of a proper
understanding of power corrections was recognized\cite{polits}.  
Subsequently, leading ${ 1 \over Q^2}$ corrections to the unpolarized leading
twist structure function $F_2$ and the longitudinal structure function $F_L$
were analyzed by the Operator Product Expansion (OPE) \cite{jafsol} and
Feynman diagram \cite{efp} approaches. Later Qiu \cite{qiu} has given an
alternate method based on special propagators (utilizing unique features of
light-cone coordinates) to simplify the analysis. 

Power correction to $F_L$ is especially interesting since the leading (twist
two) contribution to $F_L$ is perturbative in origin in contrast to the case
of $F_2$. Thus the first, non-perturbative contributions to $F_L$ occur at
${ 1 \over Q^2}$ order. The complexity of the problem of higher twist
appears in the OPE analysis which utilizes a
collinear basis, since at twist four, there appears a proliferation of operator
structures. In the Feynman diagram approach it has been shown, using a
transverse basis, that contact could be made with light-front current
algebra analysis with the result that the twist four part of $F_L$ is
given by the Fourier transform of the hadron matrix element of the minus
component of the bilocal vector current. Since the minus component of the
current involves the constrained fermion field, the relevant operator has
explicit dependence on the interaction in contrast to the well-known result
for the leading twist contribution to $F_2$ which involves the plus
component of the bilocal current. Even after many years of investigation, an
intuitive physical understanding of the interaction dependence in the
structure of $F_L$ has been elusive. We provide herein a physically intuitive
picture.

Another important problem of current interest is the perturbative aspects of
the twist four matrix element. Simple power counting indicates that in the
bare theory the twist four matrix element will be afflicted with quadratic
divergences \cite{ji}. Understanding the origin and the nature of these divergences
will be quite helpful in finding procedures to remove them ( the process of
renormalization).

A third motivation to study the twist
four part of $F_L$ comes from the present status of deep inelastic
scattering experiments. Measurements\cite{exp} of the ratio of longitudinal to
transverse cross section in unpolarized deep inelastic scattering show
\cite{mira} that
power corrections play an important role in nucleon structure experiments in
the SLAC Kinematic range. It is important to go beyond phenomenological
parameterizations for a proper understanding of the non-perturbative nature
of these corrections.         

Light-front analysis of deep inelastic scattering provides an intuitive
physical picture of various structure functions at the twist two level.
Recently, the resolution of an ambiguity at the operator level and the
parton interpretation of the transverse component of the bilocal current
have been achieved in an approach based on light-front field theory\cite{hari1}.   
The physical picture of the transverse polarized structure function
\cite{wh} and a
critical examination of the Wandura-Wilczek sum rule in perturbation 
theory \cite{hari2} have also been provided in the same approach. Both
nonperturbative and perturbative issues can be addressed in the same language
in this formalism which uses the Fock space expansion for all the operators and
multiparton wave functions for the state \cite{hari96}. The approach 
also provides insights into various renormalization issues associated with the 
different components of currents and the Hamiltonian.

In this work, we show that using the same framework, one can resolve 
outstanding issues associated with the twist four
contributions to the longitudinal structure function. A brief summary of
some of our results is presented in Ref. \cite{hari3}. 
In this work we extend our previous calculations and also 
present several new results. 
Our starting point is the Bjorken-Johnson-Low (BJL)
expansion for the forward virtual photon-hadron Compton scattering
amplitude. This leads us to the commutator of currents which we present in
detail for arbitrary flavors in SU(3). Next we consider the specific case of
electromagnetic currents and arrive at expressions for the twist two part of
$F_2$ and twist four part of $F_L$ in terms of specific flavor 
dependent form factors.
In the rest of the paper we consider the flavor singlet part of the
structure functions.
We identify the integral of ${F_L(x) \over x}$ with
the fermionic part of the light-front QCD Hamiltonian density. The
consideration of mixing in the flavor singlet channel
leads us to the the definition of the twist four longitudinal gluon
structure function and then we find a sum rule, free from radiative
corrections. 

The sum rule which the physical structure function has to satisfy involves the
physical mass of the hadron which is a finite quantity. A theoretical
evaluation of the sum rule which starts with the bare theory, on the other 
hand, will be afflicted with various divergences (see Sec. IV.) 
depending on the regulator
employed. In order to compare with the physical answer resulting from the
measurement, we need to renormalize the result by adding counterterms. For
the dressed parton target, for example, these counterterms are dictated
by
mass counterterms in the light-front Hamiltonian perturbation theory. For a
dressed gluon target, calculations in Sec. IV B show that quadratic
divergences are generated and one does not automatically get the result
expected for a massless target. The divergence generated is shown to be
directly related to the gluon mass shift in old fashioned perturbation
theory. To a given order in perturbation theory, counterterms have to be
added to the calculated structure function. The precise selection of
counterterms is dictated entirely
 by the regularization and renormalization of the
light-front QCD Hamiltonian. The choice of counterterms in the
Hamiltonian, in turn, determines the counterterms to be added to the
longitudinal structure function which results in a theoretical prediction of
the physical longitudinal structure function. Recall that in Hamiltonian
perturbation theory we cannot automatically generate a massless gluon
by clever choice of regulators. The point we emphasize is that the
twist four longitudinal structure function is one to one related to the
Hamiltonian density and that there is no arbitrary freedom in this
relationship.  

We also note that in the pre-QCD era, there have been discussions about a
possible $\delta (x)$ function contribution to the longitudinal structure function
which may appear to invalidate the sum rule 
derived ignoring such subtleties. In
two-dimensional QCD Burkardt has shown \cite{bur} that ${F_L \over x^2}$ has a
delta function contribution and he has discussed implications of this for
the sum rule for ${F_L \over x^2}$. Obviously, ${F_L \over x}$ will not be
affected by such a singular contribution and we show explicitly in Sec. V
that the sum rule is verified in two-dimensional QCD by virtue of the
't Hooft equation.        

To gain an understanding of the nature of quadratic divergences, we evaluate
the twist four longitudinal structure functions for quark and
gluon target each dressed through lowest order in perturbation theory. 
The sum rule allows us to relate these
divergences to quark and gluon mass corrections in QCD in
time-ordered light-front perturbation theory. We also verify the sum rule in a
non-perturbative context in two-dimensional QCD. We also present numerical
results for $F_2$ and $F_L$ structure functions in this model using 
wave functions calculated in a variational approximation. Finally we discuss
the relevance of our results for the problem of the partitioning
 of hadron masses
in QCD.

The plan of this paper is as follows. In Sec. II we derive the expressions for
the twist two structure function $F_2$ and the twist four longitudinal
structure function
$F_L$ using the BJL expansion and equal time ($x^+$) current algebra. 
The sum
rule for $F_L$ is given in Sec. III. In Sec. IV we 
evaluate $F_L$ for quark and gluon targets dressed through lowest order
in perturbation theory
and explicitly verify the sum rule. The sum rule is verified explicitly in a
non-perturbative context in two dimensional QCD in Sec. V. In this section,
to provide a qualitative picture, we also present numerical results for the
$F_2$ and $F_L$ structure functions in this model. In Sec. VI we discuss the
issue of the breakup of hadron mass in QCD in the context of our sum rule.
Discussion and conclusions are presented in Sec. VII and our notations and conventions        
are summarized in an appendix.
\section{Preliminaries}
In this section we present the expressions for structure functions for
arbitrary flavors in SU(3) which follow from the use of the
Bjorken-Johnson-Low
expansion and light-front current algebra. 
In terms of the flavor current 
$J_a^\mu(x) = {\overline \psi}(x) \gamma^\mu {\lambda_a \over 2} \psi(x)$,
the hadron tensor relevant for deep inelastic scattering is given by 
\begin{equation}
	W^{\mu \nu}_{ab} = {1\over 4\pi} \int d^4 \xi~ e^{iq \cdot \xi} 
		\langle P |[J^\mu_{a}(\xi), J^\nu_{b}(0)]|P \rangle  \, .
\end{equation}
The forward virtual photon-hadron Compton scattering amplitude is given by
\begin{eqnarray}	
	T^{\mu \nu}_{ab} &=& i \int d^4\xi e^{iq\cdot \xi} \langle P | 
		T(J^\mu_{a}(\xi) J^\nu_{b}(0)) |P \rangle.
\end{eqnarray}
We have
\begin{equation}	
	T^{\mu \nu}_{ab} (x,Q^2) = 2 \int_{-\infty}^\infty d{q'}^+
		{W^{\mu \nu}_{ab} (x',Q^2) \over {q'}^+ - q^+} ~.
\label{disp}
\end{equation} 

Using the BJL expansion \cite{jac}, we have,
\begin{equation} 
	T^{\mu \nu}_{ab} = -  {1\over q^-}
		 \int d\xi^- d^2 \xi_\bot e^{iq\cdot \xi}
	  \langle P | [J^\mu_{a}(\xi), J^\nu_{b}(0)]_{
		\xi^+=0}| P \rangle \, + ... \label{tmn}
\end{equation}
where $ ... $ represents higher order terms in the expansion which we ignore
in the following.
In the limit of large $q^-$, from Eq. (\ref{wmn}), we have
\begin{eqnarray}
W^{+-}_{ab} = {1 \over 2 } F_{L(ab)} + (P^\perp)^2 { F_{2(ab)} \over \nu}+ {P^\perp.
q^\perp \over x \nu} F_{2 (ab)}, \label{lqwmn}
\end{eqnarray}
with $ x = { -q^2 \over 2 \nu}$ and $ \nu =P.q$. 
On the other hand, from Eq. (\ref{tmn}), 
\begin{eqnarray}
Limit_{q^- \rightarrow \infty} ~~ T^{+-}_{ab} = -  {1\over q^-}
		 \int d\xi^- d^2 \xi_\bot e^{iq\cdot \xi}
	  \langle P | [J^+_{a}(\xi), J^-_{b}(0)]_{
		\xi^+=0}| P \rangle \, . \label{tpm}
\end{eqnarray}

The components of the 
flavor current $J_a^\mu(x)$
obey the equal- $x^+$ canonical commutation relation (to be specific, we
consider SU(3) of flavors)
\begin{eqnarray} 
\left [ J_a^+(x), J_b^-(y) \right ]_{x^+=y^+}  ~=~ 
 2 i f_{abc} ~{\overline \psi}(x)~ \gamma^-{ \lambda_c \over 2} \psi(x)~
\delta^2(x^\perp - y^\perp) ~ \delta(x^- - y^-) \nonumber \\
 ~~~~~- ~{ 1 \over 2} ~\partial^+_x ~\Bigg [ \epsilon(x^- - y^-)~
  \Big [i f_{abc} ~
{\cal V}_c^-(x \mid y)~ +~ i d_{abc}~ {\overline {\cal V}}_c^- (x \mid y) \Big ]
\delta^2 (x^\perp - y^\perp) \Bigg ] \nonumber \\
~ ~~~~+~ { 1 \over 2}~ i f_{abc} ~\epsilon(x^- - y^-)~ \partial^i_x~ \Bigg [ 
\delta^2(x^\perp - y^\perp)~  \Big [ {\cal V}^i_c(x \mid y)~ - ~\epsilon^{ij}
~{\overline {\cal A}}^j_c (x \mid y) \Big ] \Bigg ] \nonumber \\
~ ~~~~~+~ { 1 \over 2}~ i d_{abc}~ \epsilon(x^- - y^-) ~\partial^i_x~ \Bigg [ 
~\delta^2(x^\perp - y^\perp) ~ \Big [ {\overline {\cal V}}^i_c(x \mid y)
~ +~ \epsilon^{ij}~{\cal A}^j_c (x \mid y) \Big ] \Bigg ]. \label{cuco}
\end{eqnarray}
In deriving the above relations, use has been made of the relation
\begin{eqnarray}
\lambda_a \lambda_b = i f_{abc} \lambda_c + d_{abc} \lambda_c.
\end{eqnarray}
We have defined the bilocal currents as follows.
\begin{eqnarray}
{\cal V}^\mu_c(x \mid y) &&~= ~{ 1 \over 2} \Big [
{\overline \psi}(x)  { \lambda_c \over 2} \gamma^\mu
\psi(y) + {\overline \psi}(y) { \lambda_c \over 2} \gamma^\mu \psi(x) 
\Big ], \nonumber \\
{\overline {\cal V}}^\mu_c(x \mid y ) &&~= ~{ 1 \over 2 i} \Big [
{\overline \psi}(x)  { \lambda_c \over 2} \gamma^\mu
\psi(y) - {\overline \psi}(y) { \lambda_c \over 2} \gamma^\mu \psi(x) 
\Big ], \nonumber \\
{\cal A}^\mu_c(x \mid y) &&~ = ~{ 1 \over 2 } \Big [
{\overline \psi}(x)  { \lambda_c \over 2} \gamma^\mu \gamma^5
\psi(y) + {\overline \psi}(y) { \lambda_c \over 2} \gamma^\mu \gamma^5 \psi(x) 
\Big ], \nonumber \\
{\overline {\cal A}}^\mu_c(x \mid y) &&~= ~{ 1 \over 2 i} \Big [
{\overline \psi}(x)  { \lambda_c \over 2} \gamma^\mu \gamma^5
\psi(y) - {\overline \psi}(y) { \lambda_c \over 2} \gamma^\mu \gamma^5 \psi(x) 
\Big ]. \label{bi}
\end{eqnarray}  
Further, we introduce the bilocal form factors
\begin{eqnarray}
\langle P \mid {\cal V}^\mu_c(\xi \mid 0) \mid P \rangle && = P^\mu
V^1_c(\xi^2,
P.\xi) + \xi^\mu V^2_c(\xi^2, P.\xi) \\
\langle P \mid {\overline {\cal V}}^\mu_c(\xi \mid 0) \mid P \rangle &&
= P^\mu {\overline V}^1_c(\xi^2,
P.\xi) + \xi^\mu {\overline V}^2_c(\xi^2, P.\xi)  \label{bff}
\end{eqnarray}
>From Eqs.(\ref{tpm}) and  (\ref{cuco}),  we get,
\begin{eqnarray}
Limit_{q^- \rightarrow \infty} ~~ q^- ~ T^{+-}_{ab}  =
-2 i f_{abc} P^- \Gamma_c ~~~~~~~~~~~~~~~~~~~~~~~\nonumber \\
 + {q^+ \over 2} \int d\xi^- e^{{i \over 2} q^+ \xi^-} \epsilon(\xi^-)
  \Big [ f_{abc} ~
\langle P \mid {\cal V}_c^-(\xi \mid 0)~ \mid P \rangle +~  d_{abc}~
\langle P \mid  {\overline {\cal V}}_c^- (\xi \mid 0) \mid P \rangle \Big ]
\nonumber \\
 - {q^i \over 2} \int d\xi^- e^{{i \over 2} q^+ \xi^-} \epsilon(\xi^-)
  \Big [ f_{abc} ~
\langle P \mid {\cal V}_c^i(\xi \mid 0)~ \mid P \rangle +~  d_{abc}~
\langle P \mid  {\overline {\cal V}}_c^i (\xi \mid 0) \mid P \rangle \Big ]
\label{lqtmn}
\end{eqnarray}  
Note that matrix elements of ${\cal A}^\mu_c(x \mid y)$ do not contribute to
unpolarized scattering. 
Using the dispersion relation given in Eq.(\ref{disp}), together with Eqs. 
(\ref{lqwmn}) and (\ref{lqtmn}) and comparing the coefficient of $q^i$ on
both sides, we get
\begin{eqnarray}
{F_{2(ab)}(x) \over x} = { i \over 4 \pi} \int d \eta e^{-i \eta x} \Big [
f_{abc} V_c^1(\eta) + d_{abc} {\overline V}_c^1(\eta) \Big ].
\end{eqnarray}
Comparing the coefficients of $q^+$ on both sides, we get,
\begin{eqnarray}
F_{L(ab)}(x) && = { 1 \over Q^2} { i \over \pi} 
{ (q^+)^2 \over P^+} \int d \eta e^{ -
i \eta x} \Big [ f_{abc} \langle P \mid
{\cal V}_c^-(\xi \mid 0)\mid P \rangle + 
d_{abc} \langle P \mid {\overline {\cal V}}_c^-
(\xi \mid 0) \mid P \rangle \Big ]  \nonumber \\
&&  - {(P^\perp)^2 \over Q^2} { i \over \pi P^+} x^2
\int d\eta e^{-i \eta x} \Big [ f_{abc} \langle P \mid {\cal V}_c^+(\xi
\mid 0) \mid P \rangle + d_{abc} \langle P \mid {\overline
{\cal V}}_c^+(\xi \mid 0) \mid P \rangle\Big ].
\end{eqnarray}
We have introduced $ \eta = { 1 \over 2} P^+ \xi^-$.

Note that our result for $F_L$ differs from the one given in the literature
\cite{cjt}. The difference can be traced to the expression for $F_L$ that
one employs. It is customary \cite{cjt,efp} to ignore target mass $M^2$ in
the expression for $F_L$ (see Eq. (\ref{fpara})). This leads to an incorrect
expression for $F_L$ which in turn will lead to an incorrect sum rule (see
the following section).

The electromagnetic current 
\begin{equation}
J^{\mu}(x) = J^\mu_3(x) + {1 \over \sqrt{3}} J^\mu_8(x).
\end{equation} 
>From the flavor structure of electromagnetic current, we observe that, only
$d_{abc}$ contributes to the structure functions in deep-inelastic 
electron-hadron scattering. Explicitly, we have, 
\begin{eqnarray}
{F_2(x) \over x} = { i \over 2 \pi P^+} \int d \eta e^{ - i \eta x} 
  \langle P \mid {\overline {\cal V}}^+( \xi \mid 0) \mid P \rangle.
\end{eqnarray}

The longitudinal structure function is given by 
\begin{eqnarray}
F_{L}(x) && = { 2 \over Q^2} { i \over \pi} 
{ (q^+)^2 \over P^+} \int d \eta e^{ -
i \eta x} \langle P \mid {\overline {\cal V}}^-(\xi \mid 0) \mid P \rangle  
\nonumber \\
&& \qquad \qquad \qquad - 2 {(P^\perp)^2 \over Q^2} { i \over \pi P^+} x^2
\int d\eta e^{-i \eta x}  \langle P \mid {\overline {\cal V}}^+ (\xi \mid 0)
\mid P \rangle  .
\end{eqnarray} 

We have defined the functions

\begin{eqnarray}
{\overline {\cal V}}^\pm (\xi \mid 0)=
\left({2 \over 3}\right)^{3 \over 2}{\overline {\cal V}}^\pm_0(\xi \mid 0) + 
{ 1 \over 3}
{\overline {\cal V}}^\pm_3(\xi \mid 0) +
 { 1 \over 3 \sqrt{3}} 
{\overline {\cal V}}^\pm_8 (\xi \mid 0).
\end{eqnarray}

In arriving at our final results
we have used explicit values of the structure constants 
of $SU(3)$,
\begin{eqnarray}
d_{338} = { 1 \over \sqrt{3}}, ~~ d_{888} = - { 1 \over \sqrt{3}}, ~~
d_{330} = d_{880} = \sqrt{2 \over 3}.
\end{eqnarray}
${\overline {\cal V}}^\mu_0$ is the flavor singlet component of the fermion 
bilocal vector current. 
\section{Sum rule}
Consider the flavor singlet 
part of the structure functions $F_{2(f)}$ and $F_{L(f)}$ defined by 
\begin{eqnarray}
{F_{2(f)}(x) \over x} = { 1 \over 4  \pi P^+} \int d \eta e^{ - i \eta x} 
  \langle P \mid   \Big [
{\overline \psi}(\xi)   \gamma^+
\psi(0) - {\overline \psi}(0) \gamma^+ \psi(\xi) 
\Big ] \mid P \rangle. \label{f21}
\end{eqnarray}
\begin{eqnarray}
F_{L(f)}(x) && = { 1 \over Q^2} { 1 \over \pi} 
{ (q^+)^2 \over P^+} \int d \eta e^{ -
i \eta x} \langle P \mid   \Big [
{\overline \psi}(\xi)   \gamma^-
\psi(0) - {\overline \psi}(0)  \gamma^- \psi(\xi) 
\Big ] \mid P \rangle  
\nonumber \\
&& \qquad \qquad \qquad - {(P^\perp)^2 \over Q^2} { 1 \over \pi P^+} x^2
\int d\eta e^{-i \eta x}  \langle P \mid 
  \Big [
{\overline \psi}(\xi)   \gamma^+
\psi(0) - {\overline \psi}(0)  \gamma^+ \psi(\xi) 
\Big ]
\mid P \rangle . \label{flBJL1}
\end{eqnarray} 

>From Eq. (\ref{f21}) it follows that $ F_{2(f)}(-x) = F_{2(f)}(x)$ and 
from Eq. (\ref{flBJL1}) we explicitly find that $
F_{L(f)}^{\tau=4}(-x) = -F^{\tau=4}_{L(f)}(x) $. 
It can be verified \cite{hari3} that $F_{L(f)}^{\tau=4}$ satisfies the sum rule,
\begin{eqnarray}
\int_{0}^{1} dx {F_{L(f)}^{\tau=4}(x,Q^2) \over x} = { 2 \over Q^2} \Big [
\langle P \mid \theta_q^{+-}(0) \mid P \rangle - {(P^\perp)^2 \over (P^+)^2} 
\langle P \mid \theta^{++}_q(0) \mid P \rangle \Big ], \label{flsr1}
\end{eqnarray}    
where $\theta^{+-}_q =  i \overline{\psi} \gamma^- \partial^+ \psi$ is the
 light-front QCD Hamiltonian density and $ \theta^{++}_q = i \overline{\psi}
 \gamma^+ \partial^+  \psi$ is the light-front longitudinal momentum density
in the light-front gauge $A^+=0$.

Here we have used the fact that the physical structure function 
vanishes for $x >1$. Neglect of $M^2$ in the expression (Eq. (\ref{fpara}))
for $F_L$ will lead to $(P^\perp)^2 + M^2$ instead of $(P^\perp)^2$ in the
above equation which would spoil the correct sum rule given below.

The integral of ${F^{\tau=4}_{L(f)} \over x}$ is therefore 
related to the hadron matrix
element of the (gauge invariant) fermionic part of the light-front 
{\it Hamiltonian density}.
This result
manifests the physical content and the non-perturbative nature of the 
twist-four part of the
longitudinal structure function.

The fermionic operator matrix elements appearing in Eq. (\ref{flsr1}) 
change with $Q^2$ as a result of the mixing of quark and gluon operators in
QCD under renormalization.    
 Analyzing the operator mixing we obtain a new 
sum rule at the twist four level \cite{hari3}. 
\begin{eqnarray}
\int_0^1 { dx \over x} F_L^{\tau=4} = 4 {M^2 \over Q^2}. \label{flsr}   
\end{eqnarray}
Where M is the invariant mass of the hadron and
 $ F_L^{\tau=4}=  F_{L(q)}^{\tau=4} +  F_{L^(g)}^{\tau=4}$, 
$ F_{L^(g)}^{\tau=4}$ is the twist four longitudinal gluon structure function
which we define as,
\begin{eqnarray}
F_{L(g)}^{\tau=4}(x) && = { 1 \over Q^2} {x P^+ \over 2 \pi} \int dy^- ~
e^{-{i \over 2} P^+ y^- x} \nonumber \\
&& ~~~~~ \Bigg [ \Big [ \langle P \mid (-) F^{+ \lambda a}(y^-) F^-_{~~\lambda a}(0) + 
{ 1 \over 4} g^{+-} F^{\lambda \sigma a} (y^-) F_{\lambda \sigma a}(0) \mid
P \rangle + (y^- \leftrightarrow 0 ) \Big ] \nonumber \\
&& ~~~~~~~  -{(P^\perp)^2 \over (P^+)^2} \Big [ 
\langle P \mid (-)F^{+ \lambda a}(y^-) 
F^+_{~~\lambda a}(0) \mid P \rangle + ( y^- \leftrightarrow 0) \Big ] \Bigg
].
\end{eqnarray}
Where$ F^{\mu \lambda a} = \partial^\mu A^{\nu a} - \partial^\nu A^{\mu a}
+ g f^{abc} A^{\mu}_b A^\nu_c$.
Note that in the definition of$ F_{L(g)}^{\tau=4}(x)$ the  second term 
where the arguments of $F^{\lambda \sigma a}$ are interchanged is missing in 
Ref. \cite{hari3}.

To our knowledge, this is the first sum rule at the twist four level for deep
inelastic scattering or for QCD in general. The previously known sum rules
in deep inelastic scattering are all at the twist two level. The operators
involved are kinematical (light-front longitudinal momentum, light-front
helicity, etc.) in nature. In contrast, the sum rule we have derived
involves a dynamical operator (light-front QCD Hamiltonian) 
thus revealing a new aspect of  
the underlying non-perturbative dynamics. Our results show that the
measuremnent of the
flavor singlet part of the fermionic contributions to the twist four
longitudinal structure function in deep inelastic scattering directly reveals
the hadron expectation value of
the fermionic part of the light-front QCD Hamiltonian density in light-front
gauge.    
\vskip .2in

\section{Dressed parton calculations}
\subsection{Dressed quark with non-zero mass}

Next, we investigate the implications of Eq. (\ref{flsr1})
for quadratic divergences in $F_{L(q)}^{\tau=4}$ in perturbation theory.
We select the target to be a dressed quark 
and evaluate the structure functions 
to order $g^2$. That is, we take the state 
$ \mid P \rangle$ to be a dressed quark
consisting of bare states of a quark and a quark plus 
a gluon:
\begin{eqnarray}
\mid P, \sigma \rangle && = \phi_1 b^\dagger(P,\sigma) \mid 0 \rangle
\nonumber \\  
&& + \sum_{\sigma_1,\lambda_2} \int 
{dk_1^+ d^2k_1^\perp \over \sqrt{2 (2 \pi)^3 k_1^+}}  
\int 
{dk_2^+ d^2k_2^\perp \over \sqrt{2 (2 \pi)^3 k_2^+}}  
\sqrt{2 (2 \pi)^3 P^+} \delta^3(P-k_1-k_2) \nonumber \\
&& ~~~~~\phi_2(P,\sigma \mid k_1, \sigma_1; k_2 , \lambda_2) b^\dagger(k_1,
\sigma_1) a^\dagger(k_2, \lambda_2) \mid 0 \rangle. 
\end{eqnarray}

In the previous work \cite{hari3} we have given results for massless quark
state.
We have shown that the twist four longitudinal structure function  has 
quadratic
divergences in perturbation theory. In this section, we show that for a
massive quark, in addition to quadratic divergences, logarithmic divergences
are generated.
We have,
\begin{eqnarray}
F_L = {\cal M}_1 + {\cal M}_2
\end{eqnarray}
where
\begin{eqnarray}
{\cal M}_1 && = { 1 \over \pi Q^2 } \int dy^- e^{-{ i \over 2}P^+ y^-x} \langle P
\mid {\psi^{+}}^\dagger(y^-) \nonumber \\
&&~~~~~~~ \Big [ \alpha^\perp. \big  [ i \partial^\perp + g A^\perp(y)
\big ] + \gamma^0 m  \Big ] \Big [ 
\alpha^\perp . \big [ i \partial^\perp + g A^\perp(0) \big ] + 
\gamma^0 m \Big ] \psi^+(0) + h.c.
\mid P \rangle, \label{fl1} 
\end{eqnarray}
and 
\begin{eqnarray}
{\cal M}_2 = - 4{ (P^\perp)^2 \over Q^2}   x F_{2(q)}(x).
\label{fls2m}
\end{eqnarray}
where $\mid P \rangle$ now has a mass $M$ and $m$ is the bare quark mass.

In the case of quark contributions, the second term in the expression for
the bilocal current in Eq. (\ref{bi}) vanishes. 
First we evaluate the contribution ${\cal M}_2$ given in 
Eq. (\ref{fls2m}).
We obtain,
\begin{eqnarray}
{\cal M}_2=&&~~~ -4 {(P^\perp)^2 \over Q^2} x^2 \Big [ \delta(1-x) 
+ {g^2 \over 8
\pi^3} C_f \Big (\int d^2k_\perp  {{ 1+x^2 \over 1-x} k^2_\perp + (1-x)^3m^2
\over [m^2(1-x)^2 +k^2_\perp]^2}\nonumber\\&&~~~\quad\quad\quad\quad\quad\quad\quad 
\quad\quad- \delta(1-x) \int dyd^2k_\perp 
{{ 1+y^2 \over 1-y} k^2_\perp + (1-y)^3m^2
\over [m^2(1-y)^2 + k^2_\perp]^2}
\Big ) \Big ],\label{fl2f}
\end{eqnarray}
where $C_f = {N^2 - 1 \over 2N}$ for $SU(N)$.

Here we have presented the result without working out the transverse
integration to maintain a greater degree of transparency.

The contribution from ${\cal M}_1$ is split into four parts 
with additional contributions coming from quark mass terms
and can be written as follows.
\begin{eqnarray}
{\cal M}_1 
&& = { 1 \over \pi Q^2} \int dy^- e^{-{i \over 2} P^+ y^- x} \langle P \mid
{\psi^{+}}^\dagger(y^-)\big ( -(\partial^\perp)^2 + m^2\big)\psi^+(0) \mid P 
\rangle\nonumber \\
&& ~~~+ g{ 1 \over \pi Q^2} \int dy^- e^{-{i \over 2} P^+ y^- x} \langle P \mid
{\psi^{+}}^\dagger(y^-) (i \partial^\perp . \alpha^\perp +\gamma^0m)\alpha^\perp.
A^\perp(0) \psi^+(0) \mid P \rangle \nonumber \\
&& ~~~+ g{ 1 \over \pi Q^2} \int dy^- e^{-{i \over 2} P^+ y^- x} \langle P \mid
{\psi^{+}}^\dagger(y^-) \alpha^\perp.A^\perp(y) (i \partial^\perp .\alpha^\perp
+\gamma^0 m)\psi^+(0) \mid P \rangle \nonumber \\
&& ~~~+ g^2{ 1 \over \pi Q^2} \int dy^- e^{-{i \over 2} P^+ y^- x} \langle P \mid
{\psi^{+}}^\dagger(y^-) A^\perp(y).A^\perp(0) \psi^+(0)\mid P \rangle \\
&&\equiv {\cal M}_1^{a}+{\cal M}_1^{b}+{\cal M}_1^{c}+{\cal M}_1^{d}.
\label{mtot}
\end{eqnarray} 
Since the operators in Eq. (\ref{fl1}) are taken to be normal ordered, 
the
contribution of ${\cal M}_1^{d}$ vanishes to order $g^2$.
 
Explicit calculation leads to the diagonal Fock basis contributions  
\begin{eqnarray}
({\cal M}_1)_{diag}=
{\cal M}_1^{a} = &&4 {(P^\perp)^2 \over Q^2} x^2 \Big [ \delta(1-x) 
+ {g^2 \over 8
\pi^3} C_f \Big ( \int d^2k_\perp  {{ 1+x^2 \over 1-x} k^2_\perp + (1-x)^3m^2
\over [m^2(1-x)^2 + k^2_\perp]^2}\nonumber\\&& ~~~
\quad\quad\quad\quad\quad\quad
 - \delta(1-x) \int dyd^2k_\perp 
{{ 1+y^2 \over 1-y} k^2_\perp + (1-y)^3m^2
\over [m^2(1-y)^2 + k^2_\perp]^2} \Big )
\Big ]\nonumber\\&& ~~~\quad+
{4 m^2 \over Q^2}\delta(1-x)\Big [1- C_f {g^2 \over 8
\pi^3} \int dyd^2k_\perp  {{ 1+y^2 \over 1-y} k^2_\perp + (1-y)^3m^2
\over [m^2(1-y)^2 + k^2_\perp]^2}\Big ]\nonumber\\&& ~~~\quad +
{4 C_f \over Q^2} {g^2 \over 8
\pi^3} \int d^2k_\perp (k^2_\perp + m^2) {{ 1+x^2 \over 1-x} k^2_\perp + 
(1-x)^3m^2 \over [m^2(1-x)^2 + k^2_\perp]^2}
\end{eqnarray}.

The first term here explicitly cancels the term ${\cal M}_2$ given in
Eq. (\ref{fl2f}).

Off-diagonal contributions
\begin{eqnarray}
({\cal M}_1)_{nondiag}={\cal M}_1^{b}+{\cal M}_1^{c} =&& 
{C_f \over Q^2}{g^2 \over \pi^3} \Big [\delta(1-x) \int dyd^2k_\perp 
{m^2(1-y) \over [m^2(1-y)^2
+k^2_\perp]}\nonumber\\&&~~~\quad\quad\quad\quad\quad
 - \int d^2k_\perp {k^2_\perp +
m^2(1-x)^2 \over (1-x)[m^2(1-x)^2 + k^2_\perp]}\Big].
\end{eqnarray} 

Adding all the contributions, we have, 
\begin{eqnarray}
F^{\tau=4}_{L(q)}(x) = {4 m^2 \over Q^2} \delta(1-x) 
+ {4 C_f \over Q^2} {g^2 \over 8
\pi^3}\Big [ \int d^2k_\perp (k^2_\perp + m^2) 
~~~~~~~~~~~~~~~~~~~~~~~~~~~~~~~~~~~~ \nonumber \\
~~~ \times{{ 1+x^2 \over 1-x} k^2_\perp + 
(1-x)^3m^2 \over [m^2(1-x)^2 + k^2_\perp]^2}-\delta(1-x) 
m^2 \int dyd^2k_\perp {{ 1+y^2 \over 1-y} k^2_\perp + (1-y)^3m^2
\over [m^2(1-y)^2 + k^2_\perp]^2}\Big ] \nonumber \\
- {C_f \over Q^2}{g^2 \over \pi^3}\Big [
\int d^2k_\perp {k^2_\perp +
m^2(1-x)^2 \over (1-x)[m^2(1-x)^2 + k^2_\perp]}
-\delta(1-x) \int dyd^2k_\perp {m^2(1-y) \over [m^2(1-y)^2 + k^2_\perp]}\Big].
\label{fltotm}
\end{eqnarray}
Here we have used $M=m$, since the difference that it entails is higher
order in the coupling. Note that we are getting back the free quark answer
once we switch off the interaction. Also, the dressed
mass-less quark answer can be
easily regenerated by putting $M=m=0$. Note that the $k_\perp$-integration
now produces logarithmic divergences with the expected quadratic ones, as we
remarked earlier.

To check the sum rule explicitly, we evaluate the RHS of Eq. (\ref{flsr1}) next.
A straightforward evaluation leads to
\begin{eqnarray}
\langle P \mid \theta^{+-}_q(0) \mid P \rangle_{nondiag}  = 
-C_f{g^2 \over 2\pi^3} \int dxd^2k_\perp{k^2_\perp +
m^2(1-x)^3 \over x(1-x)}{1 \over [m^2(1-x)^2 + k^2_\perp]}
\end{eqnarray}
\begin{eqnarray} 
\langle P \mid \theta^{+-}_q(0) \mid P \rangle_{diag} - 
{(P^\perp)^2 \over (P^+)^2}
\langle P \mid \theta^{++}_{q}(0) \mid P
\rangle_{diag} =\nonumber\\~~~ 
2m^2 + 
2 C_f {g^2 \over 8
\pi^3} \int dxd^2k_\perp {k^2_\perp +(1-x) m^2 \over x} {{ 1+x^2 \over 1-x} k^2_\perp + 
(1-x)^3m^2 \over [m^2(1-x)^2 + k^2_\perp]^2}.
\end{eqnarray}
Adding the diagonal and off-diagonal contributions from the fermionic part
of the Hamiltonian density and multiplying it by ${2 \over Q^2}$ one obtains
the RHS of the sum rule. Comparing it with the integral of ${F_L^{\tau=4} 
\over x}$, where $F_L$ is given in Eq.(\ref{fltotm}), one easily sees that 
the sum rule is verified. 

To see the connection of $F_L$ with the fermionic mass shift, we calculate
the contribution of the gluonic part of the energy momentum 
tensor $\theta^{+-}$ to
the sum rule for the total $F_L$. Explicit calculation gives,
\begin{eqnarray}
\langle P \mid \theta^{+-}_g(0) \mid P \rangle_{nondiag}  = 
-C_f{g^2 \over 2\pi^3} \int dxd^2k_\perp{(1+x) k_\perp^2 \over (1-x)^2}{1
\over
[m^2(1-x)^2 + k^2_\perp]}
\end{eqnarray}
\begin{eqnarray} 
\langle P \mid \theta^{+-}_g(0) \mid P \rangle_{diag} - 
{(P^\perp)^2 \over (P^+)^2}&&
\langle P \mid \theta^{++}_g(0) \mid P
\rangle_{diag} =\nonumber\\&&~~~  
2 C_f {g^2 \over 8
\pi^3} \int dxd^2k_\perp {k^2_\perp \over (1-x)} {{ 1+x^2 
\over 1-x} k^2_\perp + 
(1-x)^3m^2 \over  [m^2(1-x)^2 + k^2_\perp]^2}.
\end{eqnarray}
Thus, we get,
\begin{eqnarray}
\langle P \mid \theta^{+-}_q(0)+\theta^{+-}_g(0) \mid P \rangle_{nondiag}
= \nonumber \\
-C_f{g^2 \over 2\pi^3} \int dxd^2k_\perp{{(1+x^2) \over 1-x}k_\perp^2 
+(1-x)^3m^2 \over x(1-x)}{1
\over
[m^2(1-x)^2 + k^2_\perp]}
\end{eqnarray}
\begin{eqnarray} 
\langle P \mid \theta^{+-}_q(0)+\theta^{+-}_g(0) \mid P \rangle_{diag} - 
{(P^\perp)^2 \over (P^+)^2}
\langle P \mid \theta^{++}+\theta^{++}_g(0) \mid P
\rangle_{diag} = \nonumber\\  
2 C_f {g^2 \over 8
\pi^3} \int dxd^2k_\perp {{(1+x^2) \over 1-x}k_\perp^2 
+(1-x)^3m^2 \over x(1-x)}{1
\over
[m^2(1-x)^2 + k^2_\perp]}
\end{eqnarray}
Adding diagonal and off-diagonal contributions, we get,
\begin{eqnarray} 
\langle P \mid \theta^{+-}(0) \mid P \rangle - 
{(P^\perp)^2 \over (P^+)^2}
\langle P \mid \theta^{++}(0) \mid P
\rangle = \nonumber\\~~~  
- C_f {g^2 \over 4
\pi^3} \int dxd^2k_\perp {{(1+x^2) \over 1-x}k_\perp^2 
+(1-x)^3m^2 \over x(1-x)}{1
\over
[m^2(1-x)^2 + k^2_\perp]}
\end{eqnarray}
 
Note that this result is connected to the full 
fermion mass shift $ \delta p_1^-$ in 
second order perturbation theory. We have (see Eq. (4.10)) in Ref.
\cite{qcd2},
\begin{eqnarray}
\delta p_1^- = - { 1 \over 2 P^+} C_f {g^2 \over 4 \pi^3} \int dx d^2 k_\perp 
{ {(1+x^2) \over 1-x} k_\perp^2 + (1-x)^3 m^2  \over x(1-x)}
{ 1 \over [m^2 (1-x)^2 + k_\perp^2]}.
\end{eqnarray}

\subsection{Dressed Gluon}
In this section we check the sum rule explicitly for a dressed gluon target.
 We consider the gluon to be composed of a bare gluon and a
quark anti-quark pair.
\begin{eqnarray}
\mid P, \sigma \rangle && = \phi_1 a^\dagger(P,\lambda) \mid 0 \rangle
\nonumber \\  
&& + \sum_{\sigma_1,\sigma_2} \int 
{dk_1^+ d^2k_1^\perp \over \sqrt{2 (2 \pi)^3 k_1^+}}  
\int 
{dk_2^+ d^2k_2^\perp \over \sqrt{2 (2 \pi)^3 k_2^+}}  
\sqrt{2 (2 \pi)^3 P^+} \delta^3(P-k_1-k_2) \nonumber \\
&& ~~~~~\phi_2(P,\sigma \mid k_1, \sigma_1; k_2 , \sigma_2) b^\dagger(k_1,
\sigma_1) d^\dagger(k_2, \sigma_2) \mid 0 \rangle.
\label{glu} 
\end{eqnarray}
 
The target gluon and the bare quark and anti-quark masses are taken to be
zero. Note that, to the order $g^2$, there will be a contribution from
the two-gluon Fock sector due to the non-abelian nature of the gauge coupling.
For simplicity, we exclude that contribution. It is easy to
incorporate that contribution by 
trivially extending our calculation presented here.

$F_L$ can be written in terms of ${\cal M}_1$ and ${\cal M}_2$ given 
in Eqs. (\ref{fl1}-\ref{fls2m}),
where $\mid P \rangle$ now stands for the dressed gluon represented by
Eq. (\ref{glu}). Explicit calculation gives,
\begin{eqnarray}
{\cal M}_2=&&~ -4 {(P^\perp)^2 \over Q^2} x F_{2(q)}^{dressed-gluon}\nonumber\\
=&&~ - {x^2(P^\perp)^2 \over Q^2} {g^2 \over \pi^2}N_f T_f 
\big[ x^2 + (1-x)^2 \big] ln \Lambda^2 .
\end{eqnarray}
Here $T_f={1 \over 2}$ and $N_f$ is the number of flavors.

${\cal M}_1$ is again divided into four parts as in Eq. (\ref{mtot}) and explicit
calculation in this case gives the following.
\begin{eqnarray}
{\cal M}_{1(diag)} ={\cal M}_{1(a)}=&&{x^2(P^\perp)^2 \over Q^2} 
{g^2 \over \pi^2}N_f T_f 
\big[ x^2 + (1-x)^2 \big] ln \Lambda^2 \nonumber\\&&~~~
+{\Lambda^2 \over Q^2}{g^2\over \pi^2}N_f 
T_f \big[ x^2 + (1-x)^2 \big]\label{gl1}
\end{eqnarray}
\begin{eqnarray}
{\cal M}_{1(off-diag)}=&&{\cal M}_{1(b)}+{\cal M}_{1(c)}\nonumber\\
=&&-{\Lambda^2 \over Q^2}{g^2\over \pi^2}N_f T_f 2 (1-x)
\label{gl2}
\end{eqnarray}
Thus, we get,
\begin{eqnarray}
F_L={\Lambda^2 \over Q^2} N_fT_f { g^2 \over \pi^2} \big[x^2 +(1-x)^2
-2(1-x) \big]
\label{flgl}
\end{eqnarray}
On the other hand, we get,
\begin{eqnarray} 
\langle P \mid \theta^{+-}_q(0) \mid P \rangle_{diag} - 
{(P^\perp)^2 \over (P^+)^2}
\langle P \mid \theta^{++}_{q}(0) \mid P \rangle_{diag} =
\Lambda^2 N_f T_f {g^2 \over 4\pi^2}\int dx \big[ {x^2 +(1-x)^2 \over
x(1-x)}\big]
\label{thd}
\end{eqnarray}
and
\begin{eqnarray}
\langle P \mid \theta^{+-}_q(0) \mid P \rangle_{off-diag} = 
- \Lambda^2 N_f T_f {g^2 \over 2\pi^2}\int dx \big[ {x^2 +(1-x)^2 \over
x(1-x)}\big]
\label{thod}
\end{eqnarray}
Adding diagonal and off-diagonal contributions, we get,
\begin{eqnarray} 
\langle P \mid \theta^{+-}_q(0) \mid P \rangle - 
{(P^\perp)^2 \over (P^+)^2}
\langle P \mid \theta^{++}_{q}(0) \mid P \rangle =-
\Lambda^2 N_f T_f {g^2 \over 4\pi^2}\int dx \big[ {x^2 +(1-x)^2 \over
x(1-x)}\big]
\label{thgl}.
\end{eqnarray}

Note that this
 result is connected
to the gluonic mass shift $\delta q_2^-$ due to pair production, since the
contribution from the gluonic part of the energy-momentum tensor
$\theta^{+-}_g$ in this case vanishes. 
In the massless limit, we have (see Eq. (4.40) in Ref.\cite{qcd2}),
\begin{equation}
\delta q_2^- = -{1 \over 2P^+} 
\Lambda^2 N_fT_f{g^2 \over 4\pi^2}\int dx \big[ {x^2 +(1-x)^2 \over 
x(1-x)}\big].
\end{equation}
>From Eq. (\ref{flgl}) we compute $\int dx {F_L \over x}$. Since
$x$-integration is from $0$ to $1$, it can be written in the following form.
\begin{eqnarray}
\int dx {F_L \over x} =- 
{\Lambda^2 \over Q^2} N_fT_f{g^2 \over 2\pi^2}\int dx \big[ {x^2 +(1-x)^2 \over 
x(1-x)}\big]\label{flint}
\end{eqnarray}
Comparing  Eq. (\ref{thgl}) and Eq. (\ref{flint}), one explicitly 
verifies
the sum rule for a dressed gluon target.

As we have emphasized, in the bare theory, the twist four longitudinal
structure function is afflicted with divergences. We have to add 
counterterms to carry out the renormalization procedure so that we have
physical answers. The sum rule for the bare theory clearly shows that the
quadratic divergences generated are directly related to the gluon mass shift
in second order light-front perturbation theory arising from an intermediate
quark anti-quark pair. In order to ensure a massless gluon in second order
perturbation theory, we have to add the negative of the shift as a counterterm.
After adding the counterterm, the gluon mass shift in second order perturbation
theory is zero and the twist four longitudinal structure function for a
massless gluon becomes zero. Thus, after renormalization, the sum rule is
satisfied, with a trivial (i.e., zero) gluon longitudinal structure
function.  
 
\section{1+1 dimensional QCD: Explicit Calculations}
In this section, we turn to 
two-dimensional QCD
in order to test the sum rule given in Eq.
(\ref{flsr}) explicitly in a non-perturbative context.
In 1+1 dimensions, in $A^+=0$ gauge, we have,
\begin{eqnarray}
\int_0^1 {dx \over x} F_{L(q)}^{\tau=4}(x) = { 2 \over Q^2} \langle P \mid 
\Big [ \theta^{+-}_q(0) + \theta^{+-}_g(0) \Big ] \mid P \rangle,
\end{eqnarray} 
with $
\theta^{+-}_q = 2 m^2 {\psi^{+}}^\dagger { 1 \over i \partial^+} \psi^+ $
and $ \theta^{+-}_g = - 4 g^2 {\psi^{+}}^\dagger T^a \psi^+ 
{1 \over (\partial^+)^2}
{\psi^{+}}^\dagger T^a \psi^+$.
We consider the standard one pair ($q \overline{q}$) approximation to the 
meson ground state. Explicit evaluations show that 
\begin{eqnarray}
{F^{\tau=4}_{L(q)} \over x} = {4 \over Q^2} \psi^*(x) 
{m^2 \over x (1-x)} \psi(x),
\end{eqnarray}
and
\begin{eqnarray}
\int_0^1 dx {F^{\tau=4}_{L(g)} \over x} =  {4 \over Q^2}(-) C_f {
g^2 \over \pi} \int_0^1 dx \int_0^1 dy \psi^*(x) 
{ \psi(y) - \psi(x) \over (x-y)^2} ,
\end{eqnarray}
where $ \psi(x) $ is the ground state wave function for the meson. Thus
\begin{eqnarray} 
\int_0^1 {dx \over x} F^{\tau=4}_{L}(x) = { 4 \over Q^2} \int_0^1 dx
\psi^*(x) \Big [ {m^2 \over x (1-x)} \psi(x)
- C_f {g^2 \over \pi} \psi^*(x) \int_0^1 dy { \psi(y) - \psi(x) \over (x-y)^2}
\Big ].
\end{eqnarray} 
By virtue of the bound state equation  ('t Hooft equation) obeyed by the
ground state wave function $\psi(x)$ for the meson 
\begin{eqnarray}
M^2 \psi(x) = {m^2 \over x (1-x)} \psi(x) - C_f {g^2 \over \pi} \int dy {
\psi(y) - \psi(x) \over (x-y)^2}   
\end{eqnarray}
together with the normalization condition $ \int_0^1 dx \psi^2(x)=1$, 
we easily verify that the twist four longitudinal structure function of the
meson obeys the sum rule
\begin{eqnarray}
\int_0^1 {dx \over x} F_L^{\tau=4} = { 2 \over Q^2} \langle P \mid
\theta^{+-}(0) \mid P \rangle = 4 {M^2 \over Q^2}.
\end{eqnarray}

In the same model, the contribution to the twist two structure function from
the fermionic constituents is given by
\begin{eqnarray}
F_{2(q)}(x) = (x+ 1-x) \psi^*(x) \psi(x)  = \psi^*(x) \psi(x).
\end{eqnarray}
Note that, since there are no partonic gluons or sea in this model, the
longitudinal momentum of the meson is carried entirely by the valence quark
and anti-quark. Thus the momentum sum rule is saturated by the fermionic
part of the longitudinal momentum density. On the other hand, light-front
energy density is shared between fermionic and gauge bosonic parts and as a
consequence the fermions carry only a fraction of the hadron mass.
This seemingly paradoxical situation further illuminates the difference
between the physical content of the $F_2$ and $ F^{\tau=4}_L$ sum rules. 

To get a quantitative picture, next, we explicitly calculate the structure
functions $F_{2(q)}(x)$ and ${F_{L(q)}(x) \over x}$ for the ground state
meson in two-dimensional QCD. We have parameterized  the ground state wave
function as $ \psi(x) = {\cal N} x^s (1-x)^s$ and determined the value of
$s$ variationally by minimizing $M^2$ for given values of $m^2$ and $g^2$. 
The factor ${\cal N}$ is determined from the normalization condition
$ \int_0^1 dx \psi^*(x) \psi(x) =1.$ The resulting structure functions are
presented in Fig. 1 for two different values of $m^2$.

Since both the quark and anti-quark have 
equal mass in the model, both structure
functions are symmetric about $x={1 \over 2}$. When the fermions are heavy
(Fig. 1(a)), the system is essentially non-relativistic and 
the structure functions are significant only near the region $x={ 1 
\over 2}$. When the fermions become lighter (Fig 1(b)),
contribution to the structure function from the end-point regions become
significant indicating substantial high momentum components in the ground
state wave function.    Note that ${ F_{L(q)} \over x}$ 
measures the fermion kinetic energy (in light-front coordinates). The
exponent $s$ in the wave function is a function of the fermion mass and
$s$ decreases as $ m$ decreases. In the massless limit, $s$ vanishes
\cite{berg} so that the wave function for the ground state becomes $ \psi(x) =
\theta(x) \theta(1-x)$. This results in a flat $F_2$ structure function.
However, because of the presence of $m^2$, $F^{\tau=4}_{L(q)}$ vanishes.
Because of an exact cancellation  between the self-energy and gluon exchange
contributions, the gluonic part of the $F^{\tau=4}_{L}$ also vanishes.
Thus the sum rule is satisfied exactly since, in the zero quark mass limit,
the ground state meson is massless in two-dimensional QCD.      
      
\section{Partition  of the Hadron Mass in QCD}
As is well-known, experiments that measure the twist-two part of the $F_2$
structure function yield information on the fraction of longitudinal momenta
carried by the charged parton  constituents of the
hadron (quarks and anti-quarks). The sum rule 
we have derived yields other useful information about
the hadron structure. Namely, our sum rule shows that experiments to measure
the twist four part of the longitudinal structure function will directly reveal the
fraction of the hadron mass carried by charged parton components of the hadron.
The light-front Hamiltonian provides theoretical insight into  this
fraction as follows.

According to our analysis, the twist four part of the longitudinal structure
function is directly related to the fermionic part of the light-front QCD
Hamiltonian density $\theta^{+-}_q$ in the gauge $A^+_a=0$.
Explicitly we have 
\begin{eqnarray} 
\theta^{+-}_q = 2 {\psi^{+}}^\dagger \Big [ \alpha^\perp.(i \partial^\perp + g A^\perp) 
+ \gamma^0 m \Big ]
{ 1 \over i \partial^+} \Big [ \alpha^\perp . (i \partial^\perp + g A^\perp)
+ \gamma^0 m \Big ] \psi^{+} .
\end{eqnarray} 
Thus we have the fermion kinetic energy contribution given by
\begin{eqnarray}
\theta^{+-}_{q(free)} = 2 {\psi^+}^\dagger \big [ - (\partial^\perp)^2 + m^2
\big ] \psi^+
\end{eqnarray}
and the interaction dependent part given by
\begin{eqnarray}
\theta^{+-}_{q(int)} && = 2 g {\psi^+}^\dagger \Big [ \alpha^\perp . A^\perp {
1 \over i \partial^+}( \alpha^\perp . i \partial^\perp + \gamma^0 m ) +
( \alpha^\perp. i \partial^\perp + \gamma^0 m) { 1 \over i \partial^+}
\alpha^\perp . A^\perp \Big ] \psi^+ \nonumber \\
&& ~~~~~~~~+ 2 g^2 {\psi^+}^\dagger \alpha^\perp. A^\perp \alpha^\perp.
A^\perp \psi^+ 
\end{eqnarray}
Note that the fermion kinetic energy constitutes only a
part of the total contribution from fermions. Any theoretical estimate of
the fermionic part of the longitudinal structure function 
 necessarily has to involve off-diagonal contributions from Fock states
differing in the number of gluons by one and two.
      
It is important to emphasize the difference between equal time and light-front 
Hamiltonians in the context of our calculations. The equal-time Hamiltonian
contains the scalar density term (${\bar \psi} \psi$) accompanying the quark
mass $m$. In contrast, the quark mass appears quadratically in the free part of
the light-front Hamiltonian. Recently the question of the partition of hadron
masses in QCD has been addressed by Ji \cite{jib} in the context of the
equal-time Hamiltonian and in terms of twist-two and twist-three
observables. In his analysis, the extraction of the fraction of the hadron
mass carried by the fermion constituents is not straightforward  because of the 
presence of the scalar density term. The hadron expectation value of the
strange quark scalar density remains unknown (experimentally). Our analysis,
however, shows that the twist four longitudinal structure function, once
extracted experimentally, directly yields the fraction of the hadron mass
carried by fermionic constituents.   
 
\section{Discussion and conclusions}

To gain physical intuition on the twist four longitudinal structure function
and to understand the occurrence of quadratic divergences and the associated
renormalization issues, we have studied the twist four longitudinal
structure function in an approach based on Fock space expansion methods in
light-front field theory.  We have identified the integral of $ { 1 \over
x}$ times the  
twist four part of the fermionic contribution to longitudinal structure function
as the hadron matrix element of the fermionic part of the light-front QCD
Hamiltonian density in the light-front gauge apart from an overall constant. 
We have tested this relation to order $g^2$ in QCD perturbation theory for
both dressed quark and gluon targets. Our result shows that quadratic and
logarithmic
divergences in the twist four longitudinal structure function are directly
related to mass corrections in the light-front theory.

By investigating the mixing of operators in the flavor singlet channel, we
have recently derived \cite{hari3} a new sum rule which involves the 
invariant mass of the hadron.
The validity of the sum rule has been explicitly checked in two dimensional
QCD ('t Hooft model). To get a qualitative picture of the twist four
structure function we have computed numerically both $F_2$ and $F_L$
structure functions in the 't Hooft model using the
ground state wave function calculated using a variational ansatz.

We have also discussed the implication of our results for the problem of
breakup of hadron masses in QCD in terms of fermionic and bosonic
constituents. We have emphasized the differences between equal-time and
light-front formulations relevant for this study.

Our results indicate that the experiments to measure the twist four 
longitudinal structure
function reveal the fraction of the hadron mass carried by 
the charged parton components.
Thus these experiments play a complementary role to 
 the longitudinal momentum and helicity
distribution information obtained at the twist two level. It is of interest
to investigate the feasibility of the direct measurement of the twist four
gluon structure function in high energy experiments. Recent work of Qiu,
Sterman and collaborators have shown that semi-inclusive single jet
production in deep inelastic scattering \cite{luo} and direct photon
production in hadron nucleus scattering \cite{guo} provide direct measurement 
of twist
four gluon matrix elements.

On the theoretical side we note that
some significant progress has been made recently in the bound state problem
in light-front QCD \cite{bsp} based on similarity renormalization group
method. 
In the near future, we plan to undertake a non-perturbative calculation 
(utilizing Fock space expansion and Hamiltonian
renormalization techniques) of the longitudinal structure function for a
meson-like bound state. Such a calculation will undoubtedly help to
check the validity of current phenomenological models \cite{mira} based on
simple assumptions \cite{simple} employed
in analyzing the data.  

Another important problem is the scale evolution of
the twist four structure function which is far more complicated 
than the twist two case. Recently we have provided a physical
picture of scale evolution of the $F_2$ structure function of a composite
system in terms of multi-parton wave functions 
in momentum space\cite{paper2}. We plan to
carry out a similar analysis of the twist four longitudinal structure
function elucidating all possible scale dependencies and their physical
interpretation.

\acknowledgements
We would like to thank Stan Brodsky and Wei-Min Zhang for many
enlightening discussions. This work was supported in part by the U.S.
Department of Energy under Grant No. DEFG02-87ER40371, Division of High
Energy and Nuclear Physics.
\appendix
\section{Summary of Notations and Conventions}
The hadron tensor relevant to unpolarized electron-hadron deep inelastic scattering
is given by 
\begin{eqnarray}
	W^{\mu \nu} &=&\Big(-g^{\mu \nu} + {q^\mu q^\nu 
		\over q^2} \Big) W_1(x,Q^2) + \Big(P^\mu - {P.q 
		\over q^2} q^\mu\Big)\Big(P^\nu -{P.q \over q^2} 
		q^\nu\Big)W_2(x,Q^2). \label{wmn}
\end{eqnarray}
The dimensionless functions
\begin{equation}
	F_L(x,Q^2)= 2 \Big [-W_1 + \big [ M^2 -{(P.q)^2 \over q^2}\big ] W_2
\Big ], \label{fpara}
\end{equation}
and 
\begin{equation}
 F_2(x,Q^2) = \nu W_2(x,Q^2)
\end{equation}
are the unpolarized structure functions.

We have defined, $-Q^2 = q^2 = q^+q^- - (q^\perp)^2$.

The light-front coordinates are defined by $ x^\pm = x^0 \pm x^3$. 

The constraint equation for the fermion field which follows from the Dirac
equation, in $A^+=0$ gauge is given by
\begin{eqnarray}
 \psi^-(z) = { 1 \over 4 i} \int dy^- \epsilon(z^- - y^-) \Big
[ \alpha^\perp . (i \partial^\perp + g A^\perp)+ 
\gamma^0 m \Big ] \psi^+(y^-),
\end{eqnarray}
where the antisymmetric step function 
\begin{eqnarray}
\epsilon(x^-) = - { i \over \pi} P \int {d \omega \over \omega} e^{{i \over
2 } \omega x^-}. 
\end{eqnarray} 

\eject
\centerline{\bf List of Figures}
\vskip .2in
\begin{enumerate}
\item Fermionic contributions to the structure functions 
$F_2(x)$ and ${F_L^{\tau=4} \over x}$ for the
ground state meson in the 't Hooft model for two different values of $m$,
the quark
mass. (a) $ m =5$, $ s = 4.96$. (b) $ m =1$, $ s = .70$. The parameter 
 $s$ appearing in the wave function is determined by a variational
calculation. We have set $C_f {g^2\over \pi}=1$.
\end{enumerate} 
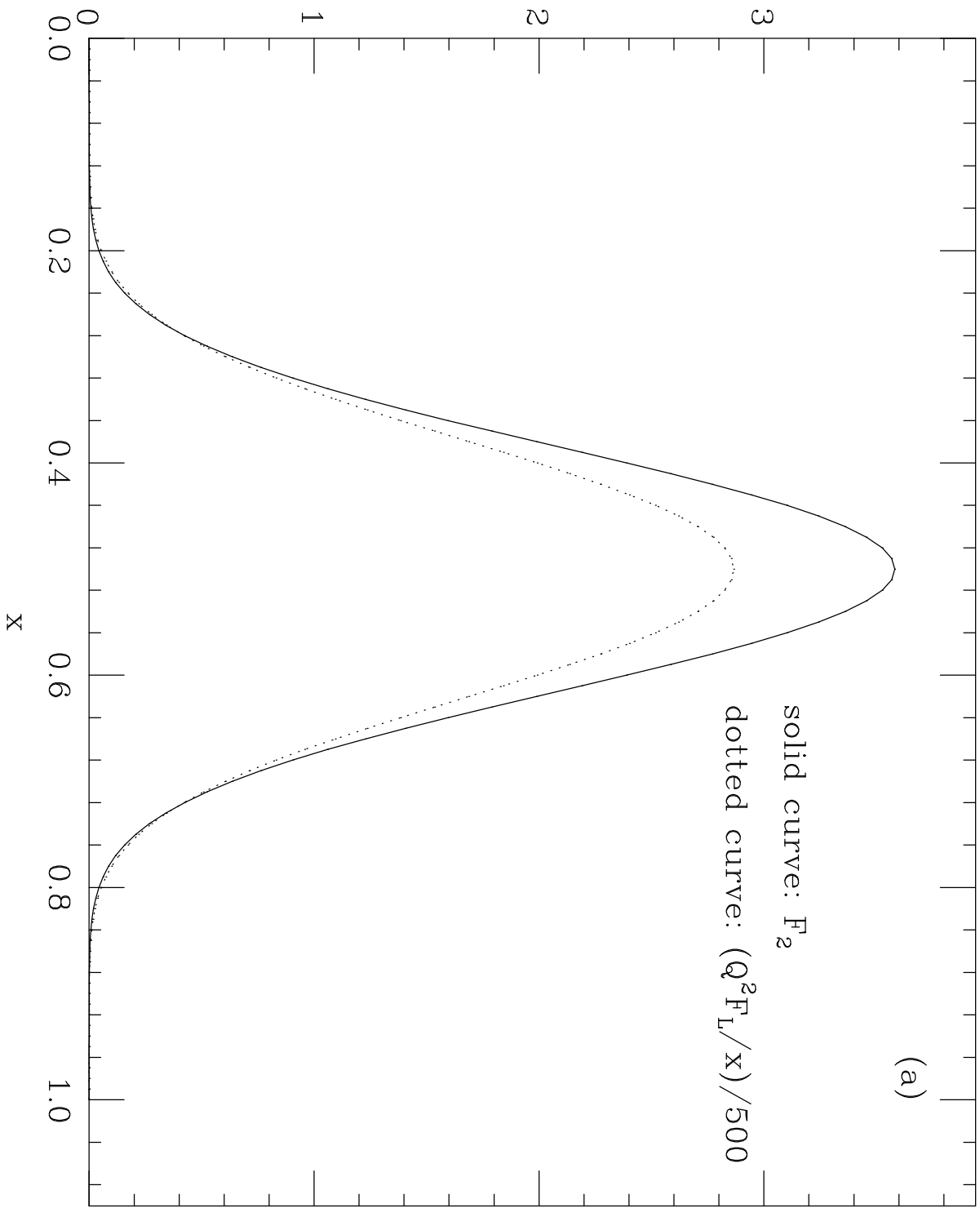
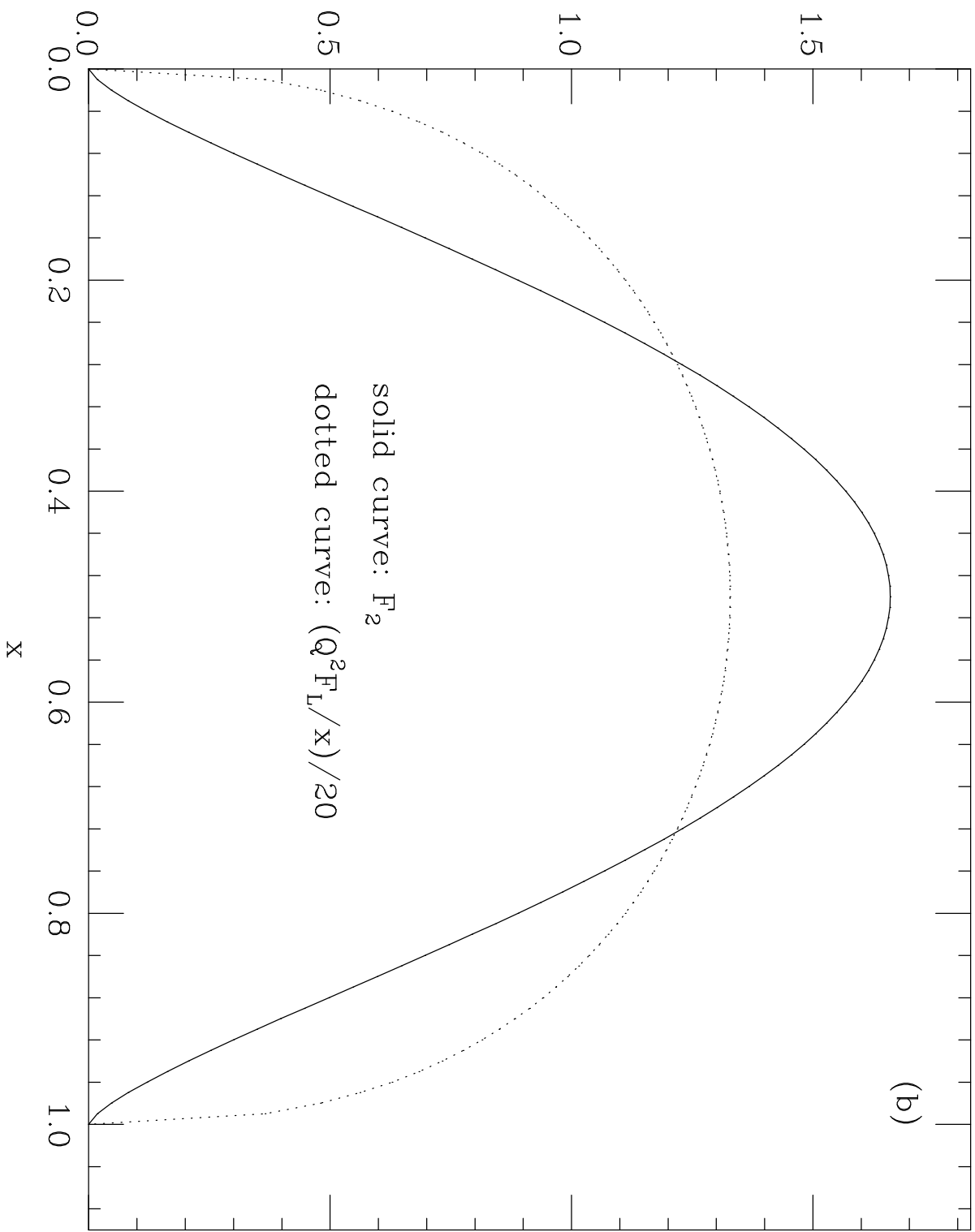
\end{document}